# Realization of Semantic Atom Blog

Dhiren R. Patel and Sidheshwar A. Khuba

**Abstract**— Web blog is used as a collaborative platform to publish and share information. The information accumulated in the blog intrinsically contains the knowledge. The knowledge shared by the community of people has intangible value proposition. The blog is viewed as a multimedia information resource available on the Internet. In a blog, information in the form of text, image, audio and video builds up exponentially. The multimedia information contained in an Atom blog does not have the capability, which is required by the software processes so that Atom blog content can be accessed, processed and reused over the Internet. This shortcoming is addressed by exploring OWL knowledge modeling, semantic annotation and semantic categorization techniques in an Atom blog sphere. By adopting these techniques, futuristic Atom blogs can be created and deployed over the Internet.

This paper ushers realization of Semantic Atom blog and discusses how to make the contents of an Atom blog human readable as well as machine accessible and processed over the Internet.

———————————— ◆ ————————————

## 1 INTRODUCTION

BLOG on the Internet is a special kind of web application. A topic of the interest is published through blog over Internet and views are solicited from the people. People having interest in the published topic participate online, post their views, state their opinions, share and publish information. Now-a-days manufacturing and marketing companies are creating blogs regarding their products & services, thus establishing the community network of consumers [1]. People publish and share information in the form of text, image, audio and video. The blogging platform provides easy means to quickly publish and share information [2]. Blog facilitates to establish a community network of the people having interest in blog topic [3]. In the blog, multimedia information builds up exponentially. In the course of time huge content is generated over blog. The analysis and reusage of content generated by the people will be very useful. For analyzing and promoting reusage of blog content it is required to be machine accessible and processed by software processes over Internet. Machine accessibility, processability and comprehending of web content are most important characteristics of semantic web [4]. The usage of semantic web technologies and techniques in blog sphere is advocated [2][5]. In this paper the strategy and techniques for semantic categorization and annotation of content published on an Atom blog is formulated and provided.

The organization of paper is as follows - in section 2 positive aspects and shortcomings of present blog have been narrated. Strategy to alleviate these short comings is addressed in section 3. The challenges identified in present blog are listed in section 4. In section 5 the gist of the widely adopted blog specifications are present.. The main characteristics of semantic blog are highlighted in section 6. The semantic information retrieval, reuse application scenario and extension of an Atom syndication format have been dealt in section 7. The paper is concluded with future plan in section 8 and with references at the end.

## 2 BLOG ASPECTS

### 2.1 Positive Aspects

**i) Multimedia Resource** - Blogging on Internet is one of the most popular activities of cyber surfers. Blogs on various topics and themes by many user communities are already there on the Internet. Also, new ones are arising. Day by day, number of bloggers and blogs on Internet are increasing. Worldwide, there are around more than 100M blogs and everyday about 50K blogs are added. The volume of information piled in blogs is exponentially increasing. In terms of total volume the huge multimedia content gets generated in web blogs. The blog is taking the form of a multimedia information resource available on the Internet. This is positive aspect of blog.

**ii) Intrinsic Knowledge** - The information published and shared over blogs is in the form of text, image, audio and video. The information published over blog is small and self contained information [6]. This is called as information snippets. The information snippet intrinsically contains knowledge. The knowledge has intangible value propositions. This is another positive aspect of blog.

### 2.2 Shortcomings

**i) Software based retrieval and reusability** - Many times it happens so that the blog entries are not noticed outside the community. The information contained in blogs is not accessible and processed by software process over Internet [7]. Because of this reason the information contents of blog can't be reused and integrated.

This is one of the major shortcomings of present day

• *Prof. Dhiren R. Patel is with Indian Institute of Technology Gandhinagar, Ahmedbad, INDIA.*
• *S.A. Khuba is with National Informatics Centre, Secretariat, Silvassa-396230, Dadra and Nagar Haveli UT, INDIA.*

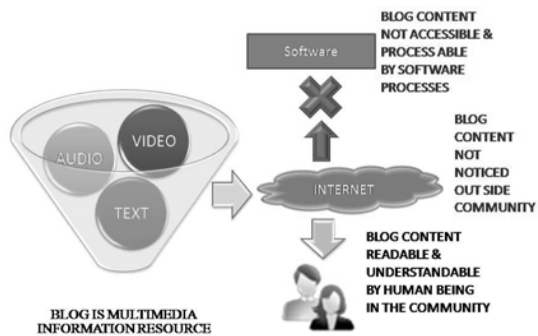

Fig. 1. Aspects of blog as information resource on Internet

blogs. The scenario of above positive aspects and short-comings in the blog is pictorially depicted in Fig. 1.

**ii) Proper Categorization** - The rise in number of blogs and exponential built of information content in blogs is posing problems in blog sphere with respect to unambiguous categorization, cataloging of blogs and navigation, search, sort, filter, querying and in aggregation of blog entry postings [6] .

## 3 STRATEGY TO ALLEVIATE SHORTCOMINGS

The shortcomings of present day blogs can be addressed by exploring application of semantic web technologies and techniques in the web blog sphere [8]. The problem of uniquely categorizing and cataloging can be addressed by tagging the blog and blog entry by the categorization term specified by internationally accepted taxonomy schema. United Nations Standard for Products and Services Code [9] is one such well known and widely accepted taxonomy schema. The OWL taxonomy of United Nations Standard for Products and Services Code is available on Internet. The problem of the software based accessibility and processing ability of the information contained in the blog can be addressed as follows - the semantic description and intrinsic knowledge contained in the contents of the blog entry posting can be abstracted, modeled and represented using OWL. The OWL semantic description and knowledge representation is recommended to be used as a metadata to annotate the contents of the blog entry posting.

## 4 CHALLENGES IN PRESENT BLOGS

In this research work, while addressing shortcomings and providing solutions following challenges in the present web blogs have been identified-

A) Where & how to specify the categorization information of the blogs and its entry postings? The specified reference categorization schema needs to be accessible over Internet, interpretable and processed by the software processes.

B) How to make data contained in blog entries accessible and processed by software processes over Internet?

C) How to computationally derive the intra relationship of blog topic with entry postings and intra relationship amongst blog entry postings?

## 5 BLOG SPECIFICATIONS

Until now plenty of syndication formats have been formulated. Most of the syndication formats are either outcome of sheer experimentation or achieving business objectives. Following three syndication formats have sustained in the hype ridden IT market. These syndication formats have been adopted on large scale.
A) Really Simple Syndication Version 2.0
B) RDF Site Summary Version 1.0
C) IETF Atom Standards Version 1.0

### 5.1 Really Simple Syndication Version 2.0

RSS is an acronym for Really Simple Syndication. RSS is a web content syndication format and XML dialect [10]. RSS originated in 1999 as a channel description framework and content-gathering mechanism. Due to its' simplicity and easy to understand form RSS became very popular, widely supported and used. RSS 2.0 specifications have been formulated and published by RSS Advisory Board. The specs formulators and publishers have confessed in the road for RSS that RSS is by no means a perfect format and specs have been frozen at version 2.0.1 [11].

### 5.2 RDF Site Summary Version 1.0

RDF Site Summary (RSS) 1.0 is a lightweight, multipurpose extensible metadata description and syndication format [12]. This is an XML application. RSS facilitates to describe better representation of relationships between metadata items. This syndication format is built upon Resource Description Format (RDF) specified by W3C.org. Due to this foundational work RDF has made it possible to incorporate intelligence into RSS 1.0 based blogging platforms.

### 5.3 IETF Atom Standards Version 1.0

The standards formulated and published by the international standards organization are backward compatible with previous releases and future releases of standards. The standards are supported by the large vendor base worldwide. Internet Engineering Task Force (IETF) is the world reputed standards body which formulates and specifies protocol in Internet domain. IETF has proposed new standards for web blogging platform. These standards are titled as Atom Publishing Protocol (APP) [13] and Atom Syndication Format [14][15]. This is a set of most recent specifications currently available for development and deployment of web blog applications.

The Atom Publishing Protocol is an application-level protocol for publishing and editing web resources using HTTP and XML 1.0. IETF Atom standards inculcate REST principals. Create – Read – Update - Delete (CRUD) operations on web resources are performed using POST – GET – PUT - DELETE methods of HTTP protocol. The Atom Syndication Format is an XML-based Web content and metadata syndication format. Atom entry document is well formed XML document. The syntax of Atom Entry Element is shown in Table 1.

TABLE 1 SYNTAX OF ATOM ENTRY ELEMENT

```
atomEntry =
    element atom:entry {
      atomCommonAttributes,
      (atomAuthor*
       & atomCategory*
       & atomContent?
       & atomContributor*
       & atomId
       & atomLink*
       & atomPublished?
       & atomRights?
       & atomSource?
       & atomSummary?
       & atomTitle
       & atomUpdated
       & extensionElement*)
}
```

IETF Atom standard has specified provision, provided facility for specifying categorization scheme and category information associated with blog entry and feed. The blog entry and feed can be categorized using "atom:category" element. The syntax of the "atom:category" element is shown in Table 2.

TABLE 2
SYNTAX OF ATOM CATEGORY ELEMENT

```
atomCategory =
    element atom:category {
    atomCommonAttributes,
    attribute term { text },
    attribute scheme { atomUri }?,
    attribute label { text }?,
    undefinedContent
    }
```

The "term" attribute is a string that identifies the category to which the entry or feed belongs. The "scheme" attribute is an IRI that identifies a categorization scheme. The "label" attribute provides a human-readable label for display in end-user applications. The first challenge in present atom blog can be addressed by making use of <atom:category/> element in blog entry, feed, collection, service and category document.

The contents of blog entry document is specified using <atom:content/> element in the atom entry xml document. A web blog application developed, deployed based on IETF Atom standards facilitates to create blog having entries in text, image, audio, video forms and specify wide variety of metadata. Atom syndication format has not specified any provision or provided facilities to annotate contents of blog entry with a semantic model. However Atom standards are extensible. There is a generic facility in the form <atom:extenssionElement /> for adding new metadata elements.

## 6 SEMANTIC BLOG

Semantic web is the future vision for present web. The content of semantic web blog will be human readable as well as software accessible and can be processed over the Internet. In Semantic blog, blog entries as well as feed are cataloged, categorized and classified by using a taxonomy term specified by internationally published taxonomy scheme [9]. The data content of blog entry is annotated with a semantic model. Semantic Model annotating blog entry will acts as one of the metadata items for the blog entry. This kind of categorization and annotation technique enables information contents of the web blogs understandable, accessible and processed by the software processes. The categorization taxonomy term and semantic metadata are used to mechanize semantic retrieval, re-usage and integration of information contents of the web blogs through software processes over Internet.

## 7 SEMANTIC INFORMATION RETRIEVAL AND REUSE

### 7.1 Application Scenario

The application scenario of semantic retrieval, integration, fusion and re-usage of multimedia content available in an Atom blog is depicted in fig.2.

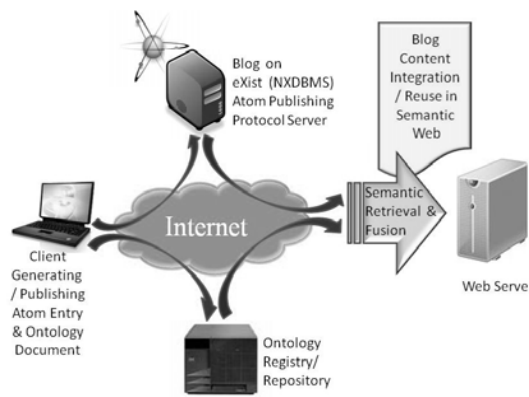

Fig. 2. Application scenario of semantic retrieval, fusion and reuse of blog content on Internet

A java client application creates an Atom entry xml document and publishes on Atom Publishing Protocol enabled web server. The semantic descriptions and knowledge intrinsically contained in the content of an atom blog entry is abstracted, modeled represented and documented as OWL file. This OWL knowledge representation and semantic modeled OWL file is required to be published on an appropriate ontology registry or repository deployed on web server [16][17]. A SOA service having capability of semantic retrieval and integration will query the OWL ontology registry and identify the IRI of those blog entry xml documents which qualifies the specified semantic similarity then auto author the new html page having hyper linkages of those identified IRIs.

### 7.2 Extension of Atom Syndication Format

Semantic blog was the research theme. The aim of research efforts was to augment the capability of multimedia content being published on an Atom blog. This capa-bility augmentation was required for software processes to access and process blog content over Internet. To accomplish this task semantic annotation technique was adopted. To enable and facilitate semantic annotation of the content it was required to have one more metadata element. To add metadata element the provision of <atom:extenssionElement /> available in Atom Syndication Format has been harnessed. Simple extension of root element viz. <atom:entry/> of an Atom entry XML document have been carried out. This extension conforms to IETF Atom Syndication Format (RFC4287). This additional metadata element facilitates semantic annotation of the data contained in <atom:content /> element by an off line OWL semantic model. The OWL semantic model acts as an additional metadata item for the multimedia content being published on an atom blog. An example of extended atom entry xml document is given in table 3.

TABLE 3
EXAMPLE OF AN EXTENDED ATOM ENTRY XML DOCUMENT

```
<?xml version="1.0" encoding="UTF-8"?>
<entry xmlns="http://www.w3.org/2005/Atom"
  xmlns:svnit=http://www.svnit.ac.in/coed/mtech/
  research/2009/khuba/ >
  <title type="text">Specifications</title>
  <id> urn:uuid:988EF5C55CDEA24EDE1251744888912</id>
  <updated>2009-08-31T18:55:12.569Z</updated>
  <author> <name>S. A. Khuba</name></author>
  <category term="45121504"
    scheme="http://www.unspsc.org/UNv1111201"
    label="Digital Camera">
  </category>
  <svnit:Semantics available="OfflineAtURL">
  http://www.daman.nic.in/khuba/ontology/camera.owl
  </svnit:Semantics>
  <contributor>
  <name>Shri. S. A. Khuba</name> </contributor>
  <content type="text">
    1) Pixels 12.3 million Effective
  .
    12) Weight is Approx. 840 g
  </content>
  <summary type="text">
    This Atom Entry XML Doc publishes tech specifications of
    Nikon D300S Digital Camera. The inline text content is
    annotated with OWL ontology named as camera.owl
  </summary>
</entry>
```

The open source software viz. Apache abdera-0.4.0-incubating.jar [18], eXist-db version 1.2.6 [19] has been used to implement and demonstrate the extension of Atom Syndication Format. Apache Abdera is a java implementation of IETF Atom Publishing Protocol (RFC 5023) and Atom Syndication Format (RFC 4287). The eXist-db is native xml database management system (NXDBMS) hosted on the web server. On eXist-db server Atom Publishing Protocol service is enabled. An ex-tended atom entry xml document is published on eXist-db server.

The metadata element named as <svnit:Semantics /> is added by harnessing <atom:extensionElement />. This element is in addition to standard metadata elements specified by Atom Syndication Formats. <svnit:Semantics /> element is used to annotate the contents of <content> element. The value of the <svnit:Semantics /> element is the IRI of owl file representing semantic description and knowledge intrinsically contained in the text content of <content /> element. The second challenge in present atom blog is addressed by making use of <svnit:Semantics /> element in blog entry document.

The line of thinking to address the third challenge is as follows – visualizing an atom blog as one family, topic of the blog as head of the family and the posted blogs as the members of the family. Describing atom blog as one owl class, it's entities as subclasses and describing inter, intra binary relationship using property owl constructs [20], then publishing the owl descriptions of an atom blog on an ontology registry or on the repository over Internet.

## 8 CONCLUSION

Exploring usage of semantic web technologies and techniques viz. OWL semantic modelling, semantic categorization and semantic annotation techniques in the Atom blog sphere is advocated. The IETF specification for Atom Syndication Format is published through RFC 4827. The semantic categorization of the Atom feed and contents published through Atom entry XML document is accomplished by usage of OWL taxonomy in <atom:Category /> element. Atom Syndication Format has not specified any provision for semantic annotation of contents published through Atom Entry XML document. This requirement has been addressed here. Atom Syndication Format has been extended to enable and facilitate off line semantic annotation of blog entry contents. The strengths of IETF Atom standards and semantic web technologies have synergized. The adoption of the formulated strategy and usage of advocated technologies and techniques augment capabilities and lays foundation for mechanizing; promoting software based semantic retrieval, re-usage, integration and fusion of the multimedia content published through an atom entry xml document over the Internet. By this way semantic Atom blog can be realized. This also facilitates to create semantically categorized and annotated multimedia content repository on NXDBMS over Internet.

Semantic web is a new research and development avenue. OWL is most recent standards available for semantic description and modeling. OWL-DL based open source semantic retrieval and integration API library is yet to be available on the Internet. The work planned to be undertaken in near future is to explore usage of OWL-DL based open source semantic software tools, similar to proprietary software MESH [21] in Atom blog sphere for semantic retrieval, integration, fusion and reuse of multimedia information over Internet.

The semantic categorization and modeling are done manually. These manual processes needs to be semi automated. Posting of entries on blog is very dynamic activity. Deriving computationally relationship between blog topic & posted blog entries, as well as relationship of blog entry postings amongst themselves is difficult task. This challenge need to be addressed in the future.

**Dr. Dhiren R. Patel** is a Professor of Computer Science & Engineering at IIT Gandhinagar, Ahmedabad, India (on leave from NIT Surat). He carries 20 years of experience in Academics, Research & Development and Secure ICT Infrastructure Design. He has played a pioneering role in bringing up Computer Engineering programs (B Tech, M Tech, Ph D) at REC/NIT Surat. His research interests cover Security and Encryption Systems, Web Services and SOA, Digital Identity Management, Low cost protocols for web based elections, and Advanced Computer Architectures. Besides numerous journal and conference articles, Prof. Dhiren has authored a book "Information Security: Theory & Practice" published by Prentice Hall of India (PHI) in 2008.

**Sidheshwar A. Khuba** is Technical Director with National Informatics Centre, Government of India, Secretariat, Silvassa, Dadra and Nagar Haveli UT, India. Promotion and implementation of various e-Governance schemes and projects is his sphere of IT activity. He is recipient of IT award instituted by Solapur Chapter of Computer Society of India. He has graduated in Electrical Engineering in 1984 from Walchand College of Engineering, Sangli in Maharashtra State, India. Presently he is pursuing postgraduate studies in IT by research at National Institute of Technology, Surat - under the supervision of Prof. Dr. Dhiren. R. Patel. His field of interest includes Web Services, SOA, Semantic Web, Semantic Grid etc.